\documentstyle[epsf,12pt]{article}
\newcommand{\ep}{\varepsilon}
\begin{document}
\begin{center}

\begin{flushright}
BI-TP 99/12
\end{flushright}

\vspace{15mm}

{\large \bf Two-loop self-energy master integrals on shell }

\vspace{15mm}

{\large
J.~Fleischer
\footnote{
E-mail: fleischer@physik.uni-bielefeld.de},~
M.~Yu.~Kalmykov
\footnote{~E-mail: misha@physik.uni-bielefeld.de \\
On leave of absence from 
BLTP, JINR, 1141980 Dubna (Moscow Region) Russia}
and A.~V.~Kotikov
\footnote{~E-mail: kotikov@sunse.jinr.ru \\
Particle Physics Lab., JINR, 1141980
Dubna (Moscow Region) Russia }
}

\vspace{15mm}

{\it ~Fakult\"at f\"ur Physik, Universit\"at Bielefeld,
D-33615 Bielefeld, Germany}

\end{center}

\begin{abstract}
Analytic results for the complete set of two-loop  self-energy master 
integrals on shell with one mass are calculated. 
\end{abstract}


\vspace*{10pt}

{\it PACS number(s)}: 12.38.Bx

\thispagestyle{empty}
\setcounter{page}0
\newpage

\begin{figure}[t]
\centerline{\vbox{\epsfysize=170mm \epsfbox{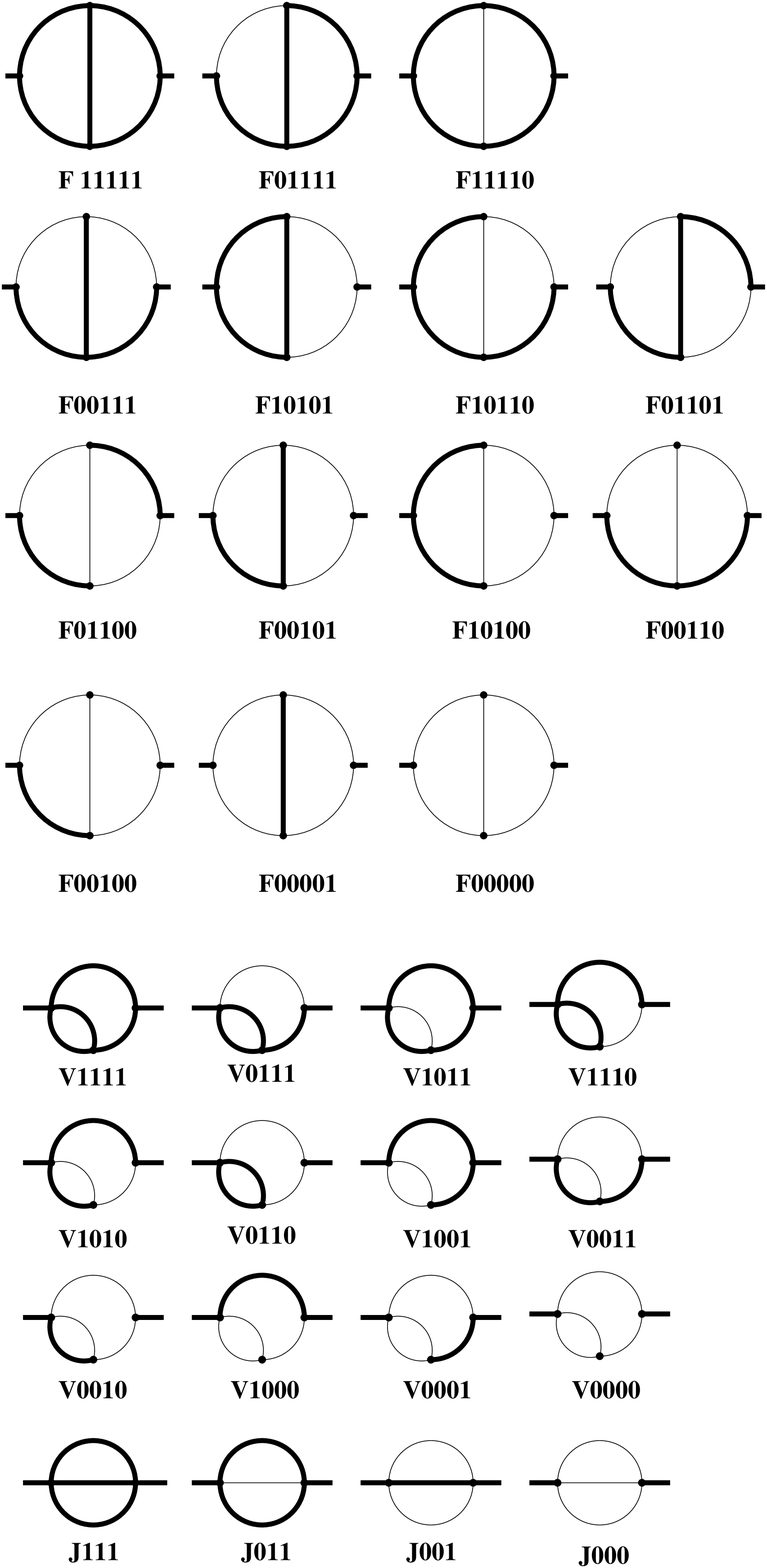}}}
\caption{\label{set} The full set of two-loop self-energies diagrams
with one mass.  Bold and thin lines correspond to the mass and
massless propagators, respectively.}
\end{figure}

For a wide class of low-energy observables the dominant and the
next-to-leading effects originate entirely from the gauge boson 
propagators. 
These corrections are sensitive to heavy virtual particles 
due to quadratic divergences of the corresponding  diagrams and 
due to non-decoupling of the heavy particles within the Higgs 
mechanism for the generation of masses.  
Only recently have the two-loop sub-leading corrections 
to the $m_W - m_Z$ interdependence of  order
$O(G_F^2 m_t^2 m_Z^2)$ been calculated in an expansion in 
terms of the large top mass \cite{subleading}. 
Moreover, the two-loop  Higgs boson mass dependence 
associated with the fermion loop
to $\Delta r$ has also become available \cite{higgs} 
\footnote{Recently the accuracy of these two results
has been tested in \cite{GSW}.}.
These  results allow to reduce the theoretical
uncertainties originating from  higher order effects of the basic 
physical quantities \cite{scheme}. 
To improve the accuracy of the
calculations, corrections of order $O(G_F^2 m_Z^4)$ must be taken into
account, which require in particular the calculation of two-loop 
self-energy diagrams with only gauge bosons and massless 
particles. 
Keeping in mind this physical application, we present 
here analytical results for all two-loop on-shell self-energy 
master integrals with one mass. 
Another application of the given 
results is the calculation of the ``naive'' part of diagrams arising 
within the asymptotic expansion when the external momentum is on the 
mass shell of the particle with the large mass \cite{asymp}. 

The aim of this letter is the analytical calculation of a full set
of two-loop on-shell self-energy master integrals with one mass, 
Eqs.(\ref{first})-(\ref{master-ons11}).   
The diagrams of this type  are presented on Fig.\ref{set}. 
It can be shown 
\cite{Tarasov1,our}, that the full set of master-integrals consists 
of the following ones: {\bf F11111, F00111, F10101, F10110, 
F01100, F00101, F10100, F00001, V1111, V1001, J111} (see Fig.\ref{set}) 
with all indices (powers of propagators)
equal to 1, and two integrals  of the {\bf J011}-type: with indices 111 
and 112, respectively (for the massless integrals see \cite{alvladim,rec}). 
The integration-by-parts method \cite{rec,rec1} 
allows not only to reduce diagrams with higher indices to
master-integrals, but also to calculate the master integrals themselves
\footnote{One of the first calculations of this type
for the Standard Model and QED was performed in Refs.\cite{S-M,1david}.}.
The main idea for the latter step is the following: a diagram with the 
index $(i+1)$ on a massive line can be represented as the derivative of 
the same diagram with  index $i$ on that line w.r.t. its mass squared.
This representation is used in certain equations obtained by the
integration by part method, yielding a differential equation
(in general partial differential equations) for the initial diagram 
(with the index $i$) which can be solved analytically. 
This approach \cite{kotikov,kotikov2}
- to construct the (differential) relations between diagrams - 
is called the `Differential Equation Method'.  
For some particular cases (where the number of the masses is small)
these partial differential equations can be rewritten as ordinary 
differential equations \footnote{An example where no reduction
to an ordinary differential equation was made, may be found in \cite{DRTJ}.}. 

To obtain the finite part of two-loop physical results one needs
to know the {\bf F}-type integrals up to the finite part, 
the {\bf V}- and {\bf J}-type integrals up to the $\varepsilon$-part, 
and the one-loop integrals up to the $\varepsilon^2$-part. 
In the last years various methods have been
suggested and many analytical results for one-,  two- and three-loop
self-energy diagrams with different values of masses and external
momenta have been obtained. 
In particular, the $\varepsilon$-part 
of one-loop integrals is given in Ref.\cite{one}, the finite parts 
of two-loop self-energy diagrams with different internal masses and 
arbitrary external momentum are given in  
Refs.\cite{1david},\cite{kotikov2},
\cite{2david}-\cite{f}, 
the $\varepsilon$-part of two-loop tadpole and sunset-type 
integrals at threshold (pseudothreshold) are calculated in
Refs.\cite{epsilon}-\cite{sunset2}.
Very efficient and elegant methods for numerical calculation
of Feynman diagrams have been presented in
Refs.\cite{Kreimer},\cite{Pivovarov}. In many of these cases the
authors have obtained one-dimensional integrals as well for arbitrary masses
and momentum. We do not know, however, of attempts to reduce these integrals to
generalized polylogarithms or other analytic expressions.

We consider here the following master-integrals in Euclidean 
space-time with dimension $N= 4-2\varepsilon$:

\begin{eqnarray}
 {\bf ONS} \{ {\cal IJ} \} (i,j,m)   & \equiv & K^{-1} 
\int d^Nk P^{(i)}(k,{\cal I}m) 
\left. P^{(j)}(k-p,{\cal J}m) \right|_{p^2=-m^2},
\nonumber \\
{\bf J} \{ {\cal IJK} \} (i,j,k,m)  & \equiv & K^{-2}
\int  d^Nk_1 d^Nk_2
P^{(i)}(k_1,{\cal I} m) P^{(j)}(k_1-k_2,{\cal J}m)
\nonumber  \\
&& ~~~~~
\times 
\left.  P^{(k)}(k_2-p,{\cal K}m) \right|_{p^2=-m^2},
\nonumber  \\
{\bf V} \{ {\cal IJKL} \}  (i,j,k,l,m)  & \equiv & K^{-2}
\int  d^Nk_1 d^Nk_2 P^{(i)}(k_2-p,{\cal I}m) 
\nonumber \\
&& 
\times 
P^{(j)}(k_1-k_2,{\cal J}m) P^{(k)}(k_1,{\cal K}m)
\left. P^{(l)}(k_2,{\cal L}m) \right|_{p^2=-m^2},
\nonumber  \\
{\bf F}\{ {\cal ABIJK} \}  (a,b,i,j,k,m)  
& \equiv & m^2 K^{-2}
\int  d^Nk_1 d^Nk_2
P^{(a)}(k_1,{\cal A} m)
\nonumber \\
&&
\times
P^{(b)}(k_2,{\cal B} m)
P^{(i)}(k_1-p,{\cal I}m)
\nonumber \\
&&
\times
P^{(j)}(k_2-p,{\cal J}m)
\left. P^{(k)}(k_1-k_2,{\cal K}m) \right|_{p^2=-m^2},
\nonumber 
\end{eqnarray}

\noindent
where $$
K = \frac{\Gamma(1+\varepsilon)}{ \left(4 \pi \right)^{\frac{N}{2}}
\left( m^2 \right)^{\varepsilon}}, ~~~ 
P^{(l)}(k,m) \equiv \frac{1}{(k^2+m^2)^l},$$ 
the normalization factor $1/(2 \pi)^N$ for each loop is assumed,
and ${\cal A,B,I,J,K} = 0,1.$

The finite part of most of the F-type master-integrals can be obtained 
from results of Ref.\cite{f} in the limit $z \rightarrow 1$. Nevertheless,
to take this limit is not an easy task in general. We have used this
approach to obtain the results for {\bf F10110} and {\bf F00101}. 
The finite part of {\bf F10101} is given in Refs.\cite{2david,analytic}, 
{\bf F00001} and {\bf F10100} 
are presented in Ref.\cite{short}. These integrals we have verified
by taking the above mentioned limit $z \rightarrow 1$ of results in Ref.\cite{f}.
{\bf F11111} is calculated in Ref.\cite{f11111} and we have checked it numerically.
Instead of the usually taken {\bf F01101} integral
\cite{2david,1david} we consider  {\bf J111} as master integral. 
We recall the results of all master integrals for completeness.
The last unknown master integral then remains to be {\bf F00111}. This
appears to be one of the most complicated integrals and below we present
details of its calculation.

The finite part of the integrals of V- and J-type can be found in 
Refs.\cite{short}-\cite{fv3}. The calculation of the $\varepsilon$ e
($\varepsilon^2$) parts of master integrals of this type have been 
performed by the differential equation method  \cite{kotikov,kotikov2}. 

The results for F-type master-integrals are follows:

\begin{equation}
{\bf F}\{ {\cal ABIJK} \} (1,1,1,1,1,m) = a_1 \zeta(3) 
+ a_2 \frac{\pi}{\sqrt{3}} S_2 + a_3 i \pi \zeta(2) + {\cal O} (\varepsilon),
\label{first}
\end{equation}

\noindent
and the coefficients $\{a_i \}$ are given in Table I:

$$
\begin{array}{|c|c|c|c|c|c|c|c|c|} \hline
\multicolumn{9}{|c|}{TABLE ~~~I}    \\   \hline
& {\bf F11111} & {\bf F00111} & {\bf F10101} & {\bf F10110} &
  {\bf F01100} & {\bf F00101} & {\bf F10100} & {\bf F00001}
\\[0.3cm] \hline
a_1 & -1 & 0 & -4 & -1 & 0 & -3 & -2 & -3 \\[0.3cm] \hline
a_2 & \frac{9}{2} & 9 & \frac{27}{2} & 9 & \frac{27}{2} &
\frac{27}{2} & 9 & 0 \\[0.3cm] \hline
a_3 & 0 & 0 & \frac{1}{3} & 0 & 1 & 1 & \frac{2}{3} & 1
\\[0.3cm] \hline
\end{array}
$$

\noindent
where \cite{S-M,2david} 
$$S_2 = \frac{4}{9\sqrt{3}} {\rm Cl}_2 \left(\frac{\pi}{3} 
\right)=0.260434137632 \cdots.$$ 
Here we used the $m^2-i \varepsilon$ prescription. The results for
the remaining master integrals are the following ones:

\begin{eqnarray}
&& {\bf V}\{ {\cal IJKL} \} (1,1,1,1,m) = 
\frac{1}{2 \varepsilon^2}
+ \frac{1}{\varepsilon}
\left( \frac{5}{2} - \frac{\pi}{\sqrt{3}} \right)
+ \frac{19}{2} + \frac{b_1}{2} \zeta(2) -4 \frac{\pi}{\sqrt{3}}
- \frac{63}{4} S_2 
\nonumber \\ && 
+ \frac{\pi}{\sqrt{3}} \ln 3
+ \varepsilon \Biggl\{ 
\frac{65}{2} + b_2 \zeta(2) - b_3 \zeta(3)
- 12 \frac{\pi}{\sqrt{3}} - 63 S_2 + b_4 \zeta(2) \ln3
+ \frac{9}{4} b_4 S_2 \frac{\pi}{\sqrt{3}}
\nonumber \\ && 
+ \frac{63}{4} S_2 \ln3
+ 4 \frac{\pi}{\sqrt{3}} \ln 3
- \frac{1}{2} \frac{\pi}{\sqrt{3}} \ln^2 3
- \frac{b_5}{2} \frac{\pi}{\sqrt{3}} \zeta (2)
- \frac{21}{2} \frac{{\rm Ls}_3 \left(\frac{2\pi}{3} \right)}{\sqrt{3}}
\Biggr\}
+ {\cal O} (\varepsilon^2),
\end{eqnarray}

\noindent
where the coefficients $\{b_i \}$ are listen in Table II:

$$
\begin{array}{|c|c|c|c|c|c|} \hline
\multicolumn{6}{|c|}{TABLE ~~~II}    \\   \hline
            &  b_1 & b_2  & b_3 & b_4 & b_5  \\[0.3cm] \hline
{\bf V1111} &  - 1 & -6 & \frac{9}{2}   & 4 & 9   \\[0.3cm] \hline
{\bf V1001} &    3 &  8 & -\frac{3}{2} & 0 & 21  \\[0.3cm] \hline
\end{array}
$$

\begin{eqnarray}
&& {\bf J111}(1,1,1,m) = - m^2 \Biggl(
\frac{3}{2\varepsilon^2} + \frac{17}{4 \varepsilon} + \frac{59}{8}
+ \varepsilon \Biggl\{\frac{65}{16} + 8 \zeta(2) \Biggr \}
\nonumber \\
&&
- \varepsilon^2 \Biggl\{
\frac{1117}{32} - 52 \zeta(2) + 48 \zeta (2) \ln 2 - 28 \zeta (3)
\Biggr\} + {\cal O} (\varepsilon^3)
\Biggr),
\label{master-j111}
\end{eqnarray}

\begin{eqnarray}
&& {\bf J011}(1,1,2,m) = \frac{1-4\varepsilon}{2 (1-2\varepsilon) (1-3\varepsilon)}
\Biggl( \frac{1}{\varepsilon^2} + 2 \frac{\pi}{\sqrt{3}}
- \frac{2}{3} \zeta(2) 
\nonumber \\
&& + \varepsilon \Biggl\{
8 \frac{\pi}{\sqrt{3}} - \frac{2}{3} \zeta(2)
-6  \frac{\pi}{\sqrt{3}} \ln 3
+ \frac{2}{3} \zeta(3)
+ 27 S_2 \Biggr\}  
+ {\cal O} (\varepsilon^2)
\Biggr),
\label{master-j011-1}
\end{eqnarray}

\begin{eqnarray}
&& {\bf J011}(1,1,1,m) = -  \frac{m^2}{2} 
\frac{4- 15 \varepsilon}{(1-2\varepsilon) (1-3\varepsilon) (2-3\varepsilon)}
\Biggl( \frac{1}{\varepsilon^2} + \frac{3}{2} \frac{\pi}{\sqrt{3}}
\nonumber \\
&& + \varepsilon \Biggl\{
\frac{45}{8} \frac{\pi}{\sqrt{3}} 
-  \frac{9}{2}  \frac{\pi}{\sqrt{3}} \ln 3
+ \frac{81}{4} S_2 \Biggr\}  
+ \varepsilon^2 \Biggl\{
12 - \zeta(2) - \frac{1863}{16} S_2 -\frac{867}{32} \frac{\pi}{\sqrt{3}} 
\nonumber \\
&& + \frac{207}{8} \frac{\pi}{\sqrt{3}} \ln 3 
+ \frac{243}{4}S_2 \ln 3
-\frac{27}{4} \frac{\pi}{\sqrt{3}} \ln^2 3
- 21 \frac{\pi}{\sqrt{3}} \zeta(2)
- \frac{81}{2}  \frac{{\rm Ls}_3 \left(\frac{2\pi}{3} \right)}{\sqrt{3}} 
\Biggr\}
\nonumber \\
&& + {\cal O} (\varepsilon^3)
\Biggr),
\label{master-j011-2}
\end{eqnarray}

\begin{eqnarray}
{\bf ONS11}(1,1,m) =
\frac{1}{1-2\varepsilon} \Biggl[
\frac{1}{\varepsilon}  - \frac{\pi}{\sqrt{3}}
+ \varepsilon \left\{ \frac{\pi}{\sqrt{3}}\ln3 - 9 S_2 \right\}
\nonumber \\
+ \varepsilon^2 \left\{
9 S_2 \ln3 - \frac{1}{2}\frac{\pi}{\sqrt{3}} \ln^2 3
- 6 \frac{{\rm Ls}_3 \left(\frac{2\pi}{3} \right)}{\sqrt{3}}
- 3 \frac{\pi}{\sqrt{3}} \zeta (2) \right\}
+ {\cal O} (\varepsilon^3) \Biggr],
\label{master-ons11}
\end{eqnarray}

\noindent
where \cite{Lewin}  
$${\rm Ls}_3 \left(\frac{2\pi}{3} \right) = -2.14476721256949 \cdots$$
(to our knowledge ${\rm Ls}_3 \left(\frac{2\pi}{3}\right)$ 
has appeared for the first time in the calculation of the 
$\varepsilon$-part of two-loop tadpole integrals in  Ref.\cite{epsilon}
\footnote{The definition is: ${\rm Ls}_3(x) = -\int_0^x \ln^2 
\left| 2 \sin \frac{\theta}{2}\right| d \theta$.} ). 
The expansion  of J111(1,1,1,m) up to $\varepsilon^4$ terms is given
in Ref.\cite{j111}. 

   The above results were checked numerically. Pad\'e approximants 
were calculated from the small momentum Taylor expansion of the
diagrams \cite{small}. The Taylor coefficients were obtained by means of the
package \cite{TLAMM} with the master integrals taken from \cite{small1}.
In this manner high precision numerical results for the on-shell
values of most of the diagrams were obtained \cite{precision}. Further we made use
of the idea of Broadhurst \cite{numeric} to apply the FORTRAN program {\bf PSLQ}
\cite{pslq} to express the obtained numerical values in terms of
a `basis' of irrational numbers, which were predicted by our analytical
calculation, i.e. the differential equation method.

Let us point out that the numbers we obtain are related to
polylogarithms at the sixth root of unity and hence are in the same class of
transcendentals obtained by Broadhurst \cite{numeric}
in his investigation of three-loop diagrams
which correspond to a closure of the two-loop topologies considered here.

   To demonstrate our method of calculation we present in what follows
some details for the {\bf F00111} integral. The basis of our approach
is the differential equation method, i.e. we write a differential equation
for {\bf F00111}, the inhomogeneous term of which is expressed in terms
of some `simpler' integrals. Before turning to the final step, we discuss
first of all these latter integrals. All calculations are performed
off shell and only at the end we take the on-shell limit.
The following extra notations are needed:

a) tadpoles:
\begin{eqnarray}
{\bf T}_1 (i,m) & \equiv & 
\int d^Nk~ P^{(i)}(k,m) =  
\frac{ \Gamma(i-\frac{N}{2}) }
{(4\pi)^\frac{N}{2}\Gamma(i)} \left( m^2 \right)^{\frac{N}{2}-i}, 
\nonumber  \\
{\bf T}_2 (i,j,k,m) & \equiv & 
\int d^Nk_1 d^Nk_2 
P^{(i)}(k_1,m) P^{(j)}(k_2,m) P^{(k)}(k_1-k_2,m), 
\nonumber  
\end{eqnarray}

b) loops with one massive line:

\begin{eqnarray}
{\bf L} (i,j,m)    \equiv  
\int d^Nk ~ P^{(i)}(k,0) P^{(j)}(k-p,m), 
\nonumber 
\end{eqnarray}

c) sunset diagrams with two massive lines:

\begin{eqnarray}
{\bf J} (k,i,j,m)  \equiv 
\int  d^Nk_1 d^Nk_2
P^{(k)}(k_2-p,0)
P^{(i)}(k_1, m)
P^{(j)}(k_1-k_2,m), 
\label{j}
\end{eqnarray}

d) two-loop diagrams with four propagators:

\begin{eqnarray}
{\bf V_1}(i,j,k,l,m)  \equiv  
\int  d^Nk_1 d^Nk_2 
P^{(i)}(k_2-p,m)
P^{(j)}(k_1-k_2,0)
P^{(k)}(k_1,m)
P^{(l)}(k_2,0), 
\nonumber \\
{\bf V_2}(i,j,k,l,m)  \equiv  
\int  d^Nk_1 d^Nk_2 
P^{(i)}(k_2-p,0)
P^{(j)}(k_1-k_2,m)
P^{(k)}(k_1,m)
P^{(l)}(k_2,m),
\nonumber
\end{eqnarray}

e) two-loop diagrams with five propagators:

\begin{eqnarray}
 {\bf F} (a,b,i,j,k,m) & \equiv &
\int  d^Nk_1 d^Nk_2
P^{(a)}(k_1,0)
P^{(b)}(k_2,0)
\nonumber \\
&&
\times
P^{(i)}(k_1-p,m)
P^{(j)}(k_2-p,m)
P^{(k)}(k_1-k_2,m).
\nonumber 
\end{eqnarray}

\noindent
{\bf 1.} {\it One-loop diagrams with one massive line.}

\noindent
Because it is ultraviolet finite, the integral with shifted indices, 
{\bf L}(1,2,m), is more suitable for analytic calculations than the `usual' 
master integral with all indices equal to one.
The recurrence relations 

\begin{eqnarray}
{\bf L} (1,1,m)\cdot(N-3) &=& -2m^2{\bf L} (1,2,m) - (p^2 + m^2){\bf L} (2,1,m),
\nonumber  \\
{\bf L} (1,1,m)\cdot(N-3) &=& {\bf T}_1 (2,m) - (p^2 + m^2){\bf L} (1,2,m),
\label{L}
\end{eqnarray}

\noindent
allow to express the needed integrals in terms of the diagram ${\bf L} (1,2,m)$,
which can be written as:

\begin{eqnarray}
{\bf L} (1,2,m) &=& 
\frac{K}{p^2}
 \int^1_0 ds
\frac{1}{(1-s)^{\ep}(1+as)^{(1+\ep)}} \nonumber \\
&=&
\frac{K}{p^2 }
\Biggl(\log{(1+a)} -\ep
\biggl[\log^2{(1+a)} + {\rm Li}_2(-a) \biggr]
+ \ep ^2 
\biggl[\frac23\log^3{(1+a)} 
\nonumber \\
&& 
+ 2\log{(1+a)}{\rm Li}_2(-a) +2{\rm S}_{1,2}(-a)
- {\rm Li}_3(-a) \biggr] \Biggr) + O(\ep ^3),
\nonumber 
\end{eqnarray}
where $a=p^2/m^2$ (a = -1 corresponds to the on-shell case). 
To obtain this result we used the representation (see e.g. \cite{f,fkv})

\begin{eqnarray}
\int d^Nk P^{(i)}(k,m_1) P^{(j)}(k-p,m_2) = 
\frac{\Gamma(i+j-N/2)}{(4\pi)^\frac{N}{2}\Gamma(i)\Gamma(j)} 
\nonumber \\ \times
\int^1_0 ds 
\frac{1}{s^{(i+1-N/2)}(1-s)^{(j+1-N/2)}}
P^{(i+j-N/2)} \left(p,\frac{m_1}{s}+ \frac{m_2}{1-s} \right).
\label{loop}
\end{eqnarray}

\noindent
{\bf 2.} {\it The sunset diagram with two massive lines.}

\noindent
The set of master-integrals of this type consists of two 
integrals, e.g. {\bf J011}(1,1,1,m) and {\bf J011}(1,1,2,m). 
To avoid ultraviolet singularities, however, we choose again
integrals with shifted indices: ${\bf J} (1,2,2,m)$ and 
${\bf J} (1,1,3,m)$. 
Appropriate recurrence relations connecting these integrals need 
to be derived (see also Refs.\cite{Tarasov1,threshold}).
Applying the integration by parts relation with the massless 
line as distinctive one
\footnote{In the integration-by-parts procedure one line is considered `distinctive',
i.e.
its propagator contains nothing but the momentum $k_{\mu}$ which is 
used for the partial integration
(see, e.g. the discussion in Ref.\cite{f}).}
, we get 

\begin{eqnarray}
&& 
{\bf J}(1,1,2,m) \cdot \left( N-2 \right) 
\nonumber \\
&&
=
2\int  d^Nk_1 d^Nk_2
P^{(1)}(k_2-p,0) 
P^{(2)}(k_1, m)
P^{(2)}(k_1-k_2,m) 
\cdot (k_2-k_1)^\mu (k_2-p)^\mu
\nonumber \\
&&
= 
\int  d^Nk_1 d^Nk_2
P^{(1)}(k_2-p,0) P^{(2)}(k_1, m) P^{(2)}(k_1-k_2,m) 
\cdot k_2^\mu (k_2-p)^\mu,
\label{sunset}
\end{eqnarray}

\noindent
where for the last step the symmetry of the integral is
taken into account. Expanding the scalar product 
$2(k_2,k_2-p)=k_2^2+(p-k_2)^2-p^2$, we have

\begin{eqnarray}
&&
{\bf J} (1,1,2,m)\cdot \left( N-2 \right) 
= \frac{1}{2}{\bf T}_1^2 (2,m) - \frac{p^2}{2} {\bf J} (1,2,2,m)
\nonumber \\
&&
+ \frac{1}{2} \int  d^Nk_1 d^Nk_2
P^{(1)}(k_2-p,0) 
P^{(2)}(k_1, m)
P^{(2)}(k_1-k_2,m) 
\cdot k_2^2.
\label{sunset1}
\end{eqnarray}

\noindent
Applying now the integration-by-parts relation with the massive 
line with index $2$ as distinctive one, we obtain

\begin{eqnarray}
&&
{\bf J} (1,1,2,m) (N-6) =
- 4m^2{\bf J} (1,1,3,m) - 2m^2 {\bf J} (1,2,2,m)
\nonumber \\
&&
-  \int  d^Nk_1 d^Nk_2
P^{(1)}(k_2-p,0) 
P^{(2)}(k_1, m)
P^{(2)}(k_1-k_2,m) 
\cdot k_2^2.
\nonumber \\
\label{sunset2}
\end{eqnarray}

\noindent
Combining Eq.(\ref{sunset1}) and Eq.(\ref{sunset2}) we get the final relation:

\begin{eqnarray}
{\bf J} (1,1,2,m) 2(1-3\ep) = 
{\bf T}_1^2 (2,m) - (p^2 + 2m^2){\bf J} (1,2,2,m)
- 4m^2{\bf J} (1,1,3,m).
\label{sun1}
\end{eqnarray}

\noindent
The second recurrence relation,  

\begin{eqnarray}
{\bf J} (1,1,1,m) \left(1 - 2 \varepsilon \right) 
\left( 1-\frac{3}{2\ep} \right) = 
\left[  \frac{\ep}{2} p^2 - 
\left( 2-  5 \ep \right) m^2 \right]{\bf J}(1,1,2,m)
\nonumber \\
- m^4{\bf J} (1,2,2,m)- m^2 (p^2 + 2m^2){\bf J} (1,1,3,m),
\label{sun3}
\end{eqnarray}

\noindent
can be obtained by applying the operator ${\Bigl(\partial/ \partial
p_{\mu}\Bigr)}^2$ to the sunset integral ${\bf J}(1,1,1,m)$, where the 
replacement $k_2 \to k_2 +p $ is performed in (\ref{j}).
Further use is made of the relation 
$${\Bigl( \partial/ \partial p_{\mu}\Bigr)}^2 P^{(i)}(p,m) = 
4i(i+1-N/2)P^{(i+1)}(p,m) - 4i(i+1)m^2 P^{(i+2)}(p,m),
$$

\noindent
and the dimension property of the integral 
(see e.g. in Ref.\cite{remiddi}): 

$$
\left( 
p^2 \frac{\partial}{\partial p^2} 
+  m^2\frac{\partial}{\partial m^2} \right) {\bf J}(i,j,k,m)
= \left( N-i-j-k \right) {\bf J}(i,j,k,m).
$$

\noindent
is taken into account.
The third recurrence relation, which we need,
has the following form: 

\begin{eqnarray}
{\bf J} (2,1,1,m) = {\bf J} (1,1,2,m) - \frac{2m^2}{\ep}
{\bf J} (1,1,3,m).
\label{sun2}
\end{eqnarray}

\noindent
Eq.(\ref{sun2}) is obtained in a similar manner as Eq.(\ref{sun3}):
the operator ${\Bigl(\partial/ \partial p_{\mu}\Bigr)}^2$ 
is applied to the sunset ${\bf J}(1,1,1,m)$ integral with the initial 
distribution of momenta (\ref{j}), and to the same diagram with the replacement 
$k_2 \to k_2 +p $. Their difference yields the recurrence relation (\ref{sun2}). 
We would like to note that Eq.(\ref{sun2}) shows that the sunset diagram
${\bf J} (1,1,3,m)$ should be known up to terms of order $\ep^2$.

After tedious calculation using representation (\ref{loop}), we obtain

\begin{eqnarray}
&& {\bf J} (1,2,2,m) 
=
\frac{K^2}{p^2}
\Biggl(\log^2{y} +2\ep
\biggl[\frac12\log^3{y}-\zeta (2)\log{y}
+\log{(1-y)}\log^2{y} 
\nonumber \\
&&+ 2\log{y} \Bigl(3{\rm Li}_2(-y)+2{\rm Li}_2(y)\Bigr)-3\zeta (3) 
-12 {\rm Li}_3(-y)-6 {\rm Li}_3(y) \biggr] \Biggr) + O(\ep ^2),
\nonumber \\
&&
{\bf J} (1,1,3,m) =
\frac{K^2}{m^2} 
\frac{1}{2 (1-2\ep)}
\Biggl(\frac{1}{\ep} + r \Biggl[\log{y} 
+\ep \Biggl\{
2\log{y}\log{\biggl(\frac{y}{(1+y)^3(1-y)}\biggr)}
\nonumber \\ && 
- \zeta (2)
-6 {\rm Li}_2(-y)-2 {\rm Li}_2(y) \Biggr\}
+\ep^2 \Biggl\{
6{\rm S}_{1,2}(y^2)-8{\rm S}_{1,2}(y)+24{\rm S}_{1,2}(-y)
\nonumber \\ && 
-8{\rm Li}_3(y)-24{\rm Li}_3(-y)-11\zeta (3) -2\zeta (2)
\log{\biggl(\frac{y^2}{(1+y)^3(1-y)}\biggr)}
\nonumber \\ && 
+4\biggl[{\rm Li}_2(y)+3 {\rm Li}_2(-y) \biggr]\log{\Bigl((1+y)^3(1-y)\Bigr)}
 +2\log{y}\log{(1-y)}\log{\Bigl(y(1-y)\Bigr)}
\nonumber \\ && 
+\frac16 \log^3{y}+\frac32\log{y}\log{\biggl(\frac{y}{(1+y)^2}\biggr)}
\log{\biggl(\frac{y}{(1+y)^6(1-y)^4}\biggr)}
\Biggr\}
\Biggr]
\nonumber \\ && 
 -\ep \frac{3a+4}{2a} \Biggl[
\log^2{y}- \ep 
\Biggl\{
12{\rm Li}_3(y)+24{\rm Li}_3(-y)+6\zeta (3) 
\nonumber \\ && 
+2\log{y}\Bigl(\zeta (2)- 4{\rm Li}_2(y)-6 {\rm Li}_2(-y) \Bigr)
-2\log^2{y}\log{(1-y)}- \log^3{y} 
\Biggr\}
\Biggr]
\Biggr) 
+ O(\ep ^3),
\nonumber 
\end{eqnarray}

\noindent
where $r^2=(4+a)/a$ and $y=(r-1)/(r+1)$.
Putting $a=-1$, we get 

\begin{eqnarray}
{\bf J011}(1,2,2,m) = \frac{2}{3 m^2}
\Biggl( \zeta(2) - \varepsilon \zeta(3)
+ {\cal O} (\varepsilon^2)
\Biggr), 
\label{j22}
\end{eqnarray}

\noindent
and 

\begin{eqnarray}
&& {\bf J011}(1,1,3,m) = \frac{1}{2 (1-2\varepsilon) m^2}
\Biggl( \frac{1}{\varepsilon} - \frac{\pi}{\sqrt{3}}
+ \varepsilon \Biggl\{
3 \frac{\pi}{\sqrt{3}} \ln 3
- \frac{\zeta(2)}{3} - \frac{27}{2} S_2 \Biggr\}
\nonumber \\
&& 
+ \varepsilon^2 \Biggl\{
\frac{ \zeta(3)}{3} 
- 27 \frac{{\rm Ls}_3 \left(\frac{2\pi}{3} \right)}{\sqrt{3}}
- 14 \frac{\pi}{\sqrt{3}} \zeta(2) 
- \frac{9}{2} \frac{\pi}{\sqrt{3}} \ln^2 3 
+ \frac{81}{2} S_2 \ln 3
\Biggr\} 
+ {\cal O} (\varepsilon^3)
\Biggr).
\label{j31} 
\end{eqnarray}

\noindent
Note that the
integrals {\bf J011}(1,1,3,m) and {\bf J011}(1,2,2,m) are much simpler than
the master integrals (\ref{master-j011-1}) and (\ref{master-j011-2}).

\noindent
{\bf 3.} {\it The {\bf F00111} diagram.}

\noindent
The  diagram {\bf F00111} demands special consideration. The investigations are 
similar to those done in \cite{f}. Applying the integration-by-parts
relation three times to one of the triangles in ${\bf F}(1,1,1,1,1,m)$ 
(every time with different distinctive line) and using
the identity

$$
2{\bf F} (1,1,2,1,1,m) + {\bf F} (1,1,1,1,2,m) = 
-\frac{d}{dm^2}  {\bf F} (1,1,1,1,1,m),
$$

\noindent
we obtain the differential equation:

\begin{eqnarray}
\biggl[(N-4)(p^4+2p^2m^2+3m^4) -2m^2(p^4+p^2m^2+m^4)
\frac{d}{dm^2} \biggr] {\bf F} (1,1,1,1,1,m)
=f,
\label{eq1}
\end{eqnarray}
where
the inhomogeneous term $f$ contains only diagrams of 
${\bf V_i}$, ${\bf J}$
and ${\bf L}$-type:

\begin{eqnarray}
&& 
f = 
(p^4-m^4) \biggl[ 
\Bigl({\bf L}(1,2,m) + {\bf L}(2,1,m)\Bigr) {\bf L}(1,1,m) 
\nonumber \\
&&
- {\bf V_1}(2,1,1,1,m) -  {\bf V_2}(2,1,1,1,m) \biggr]
\nonumber \\
&&  
+3m^2(p^2+m^2){\bf V_2}(1,2,1,1,m)-m^2(p^2-m^2){\bf V_1}(1,2,1,1,m).
\label{inh1} 
\end{eqnarray}

\noindent
Expressing all {\bf L}'s in terms of {\bf L}(1,2,m) (see (\ref{L}))
and using the recurrence relations for ${\bf V}$-type integrals
obtained via integration by parts,
we derive a new, simpler representation for the inhomogeneous 
term $f$:

\begin{eqnarray}
 f&=& \frac{(p^2-m^2)}{(1-2\ep)} 
\biggl[\frac{1-4\ep}{1-2\ep} {\bf T}_1^2(2,m)
- \frac{p^2(3-8\ep)+m^2(1-4\ep)}{(1-2\ep)} {\bf T}_1(2,m){\bf L}(1,2,m) 
\nonumber \\
&&+ 2(p^2+m^2){\bf J}(1,2,2,m) + 4m^2 {\bf J}(1,1,3,m)
+ 2m^2{\bf V_1}(1,1,2,1,m) \biggr]
\nonumber \\
&&+6m^2 \Bigl[1+ (p^2-m^2)\frac{d}{dm^2}\Bigr] m^2{\bf V_2}(1,1,2,1,m),
\label{inh2}
\end{eqnarray}

\noindent
where for the {\bf V}-type diagrams the following differential equations
hold:

\begin{eqnarray}
&&
\biggl[(N-4)(p^2+m^2)-m^2 -m^2(p^2+m^2)
\frac{d}{dm^2} \biggr] {\bf V_1} (1,1,2,1,m)
\nonumber \\
&&
= m^2 {\bf J}(1,2,2,m)
+2(p^2-m^2)\biggl[{\bf T}_1(3,m){\bf L}(1,1,m)-{\bf J}(1,1,3,m)\biggr],
\label{eq2}
\\
&&
\biggl[(N-6)p^2 +3(N-4) m^2 -2m^2(p^2+m^2)
\frac{d}{dm^2} \biggr] {\bf V_2} (1,1,2,1,m)
\nonumber \\
&&
= - p^2 {\bf J}(1,2,2,m) + 3m^2 {\bf T}_2(1,2,2,m),
\label{eq3}
\end{eqnarray}

\noindent
and (see e.g. Eq.(5.5) in Ref.\cite{small1})

$$
{\bf T}_2 (1,2,2,m)  = \frac{K^2}{m^2}\cdot 3S_2 + O(\ep).
$$

\noindent
We integrate Eqs.(\ref{eq1}), (\ref{eq2}), (\ref{eq3}) with the 
boundary condition that all diagrams tend to zero as $m^2 \to \infty$.
Finally for the on-shell case we obtain

\begin{eqnarray}
{\bf F00111}(1,1,1,1,1,m)  =  
8 \int_0^1 \frac{dx}{\sqrt{x(4-x)}} \arctan  \sqrt{\frac{x}{4-x}} 
\nonumber \\
\times
\Biggl\{ 
\frac{\sqrt{3}}{1-x} 
\left( \frac{\pi}{6} - \arctan \frac{2x-1}{\sqrt{3}} \right) 
- 
\ln \left( 1 -x +x^2\right) 
\Biggr \}.
\nonumber 
\end{eqnarray}

\noindent
This integral is evaluated numerically with a precision of 40 decimals.
Then the program {\bf PSLQ} (see above) is applied with 15 basis elements
occurring in (\ref{first}) to (\ref{master-ons11}) with the
result given in Table I.

\noindent
{\bf Acknowledgments}
We are grateful to A.~Davydychev, D.~Kreimer and O.~Veretin
for useful discussions and careful reading of the manuscript
and to O.~Veretin for his help in numerical calculation.
Two of us (J.F. and M.K.) are very indebted to O.~V.~Tarasov for 
useful comments. M.K. and A.K.'s research has been supported by the DFG 
project FL241/4-1  and in part by RFBR $\#$98-02-16923.

\end{document}